\documentclass[runningheads,11pt]{llncs}
\usepackage{graphicx}
\usepackage{multirow}
\usepackage{booktabs}
\graphicspath{ {./images/} }
\usepackage{url} 
\usepackage{tikz}
\usetikzlibrary{intersections}
\usetikzlibrary{arrows.meta}
\usetikzlibrary{positioning}
\usetikzlibrary{calc}
\usetikzlibrary{fit,shapes.arrows}
\usetikzlibrary{decorations.pathreplacing}
\definecolor{darkyellow}{rgb}{0.6, 0.5, 0}
\definecolor{skyblue}{rgb}{0.53, 0.81, 0.92}
\definecolor{myviolet1}{RGB}{215, 142, 250}

\usepackage[ruled,linesnumbered,boxed,commentsnumbered,longend]{algorithm2e}
\usepackage[hidelinks]{hyperref}
\usepackage{fullpage}
\usepackage{threeparttable}
\usepackage{tabularx}
\usepackage{array}
\usepackage{xspace}
\usepackage[colorinlistoftodos,prependcaption,textsize=tiny,textwidth=\marginparwidth,tickmarkheight=.5em]{todonotes}
\usepackage{enumitem}
\usepackage{layout} 
\newcolumntype{L}{>{\raggedright\arraybackslash}p{0.2\linewidth}}
\newcolumntype{B}{>{\raggedright\arraybackslash}p{0.25\linewidth}}
\newcolumntype{N}{>{\centering\arraybackslash}p{0.0875\linewidth}}
\newcolumntype{M}{>{\centering\arraybackslash}p{0.15\linewidth}}
\newcommand{\tool}{{\sc LCPan}\xspace}

\newcommand{\api}{{\sc Lcptools}\xspace}
\newcommand{\vgtool}{\texttt{vg}\xspace}
\newcommand{\graphaligner}{\texttt{GraphAligner}\xspace}

\newcommand{\pbsv}{\texttt{pbsv}\xspace}

\newcommand*\circled[1]{\tikz[baseline=(char.base)]{%
            \node[circle, fill=black, text=white,inner sep=1.2pt] (char) {#1};}}

\newcommand{\revise}[1]{{\color{black} #1}}

\newcommand{\junk}[1]{}

\begin{document}

\newcommand{\TableMinHash}{
    \begin{table*}[htbp]
        \centering
        \caption{The experimental results of MinHash in CHM13v2.0}
        \label{tab:minhash-fasta}
        \begin{threeparttable}
            \begin{tabular}{@{}lrrrrr@{}}
                \toprule
                \textbf{Metric} & ~~\textbf{-} & ~~\textbf{-} & ~~\textbf{-} & ~~\textbf{-} & \textbf{-} \\ \midrule
                K-mer size\dag & 31 & 31 & 31 & 31 & 31 \\
                Total k-mers & 1,053 & 5,557 & 62,865 & 654,715 & 6,859,574 \\
                Unique k-mers & 1,000 & 5,000 & 50,000 & 500,000 & 5,000,000 \\
                Avg Dist. & 2,772,382.44 & 555,777.18 & 49,545.28 & 4,761.02 & 454.42 \\
                StdDev Dist. & 3,180,719.52 & 841,584.81 & 96,152.02 & 7,712.57 & 463.73 \\
                \bottomrule
            \end{tabular}

            \begin{tablenotes}
                {\footnotesize The MinHash calculations are done using the Mash tool.}
            \end{tablenotes}
        \end{threeparttable}
    \end{table*}
}

\newcommand{\TableLcpCHM}{
    \begin{table*}[htbp]
        \centering
        \caption{The experimental results of LCP on CHM13v2.0}
        \label{tab:lcp-fasta-dct1}
        \begin{threeparttable}
            \begin{tabular}{@{}lrrrrrrrr@{}}
                \toprule
                \textbf{LCP Level} & \textbf{1} & \textbf{2} & \textbf{3} & \textbf{4} & \textbf{5} & \textbf{6} & \textbf{7} & \textbf{8}  \\ \midrule                
                Total \# of Cores & 1,385,807,259 & 595,674,177 & 253,384,737 & 108,531,321 & 46,426,965 & 19,885,308 & 8,514,640 & 3,646,315 \\
                Unique Cores & 1,395 & 517,688 & 159,701,016 & 93,601,379 & 42,859,941 & 18,795,370 & 8,191,525 & 3,563,854 \\
                Exec. Time (sec) & 71.94 & 68.72 & 31.31 & 13.65 & 5.68 & 2.48 & 1.04 & 0.44 \\
                \midrule
                Avg Distance & 2.25 & 5.23 & 12.30 & 28.72 & 67.14 & 156.76 & 366.10 & 854.87 \\
                StdDev Distance & 1.04 & 1.85 & 4.21 & 8.74 & 19.95 & 45.24 & 104.24 & 241.08 \\
                \midrule
                Avg Length & 3.67 & 10.47 & 26.58 & 63.84 & 151.23 & 354.89 & 830.88 & 1,941.85 \\
                StdDev Length & 1.01 & 1.90 & 4.32 & 8.83 & 19.49 & 43.02 & 97.57 & 222.96 \\
                \midrule
                Decrease in Core Count & 0.44 & 0.43 & 0.43 & 0.43 & 0.43 & 0.43 & 0.43 & 0.43 \\
                Increase in Avg Length & 3.67 & 2.85 & 2.54 & 2.40 & 2.37 & 2.35 & 2.34 & 2.34 \\
                Increase in Avg Distance & 2.25 & 2.33 & 2.35 & 2.33 & 2.34 & 2.33 & 2.34 & 2.34 \\
                \midrule
                Total Size (GB) & 56.79 & 24.41 & 10.38 & 4.45 & 1.90 & 0.81 & 0.35 & 0.15 \\
                \bottomrule
            \end{tabular}
            
            \begin{tablenotes}
                \footnotesize
                \item We used \api to generate LCP statistics using the CHM13v2.0 genome with a single iteration of DCT. 
                Decrease in Core Cnt. is the proportional reduction in the number of cores compared to the previous level. 
                Increase in Avg Len. is the proportional increase in the average core length compared to the previous level. 
                Increase in  Avg Dist is the proportional increase in the average distance between start positions of consecutive cores compared to the previous level. 
                Notably, while the decrease in the core count is consistently $0.43$, the increase in the average core length and the increase in the average distance converge to $1/0.43=2.33$. 
            \end{tablenotes}
        
        \end{threeparttable}
    \end{table*}
}

\newcommand{\TableMinSyncUhsCHM}{
    \begin{table*}[htbp]
        \centering
        \caption{Properties of Minimizers, Syncmers, and UHS compared to LCP in CHM13v2.0. }
        \label{tab:min-sync-uhs-fasta}  
        \begin{threeparttable}
            \begin{tabular}{@{}lrrrrrrr@{}}
                \toprule
                \textbf{Metric} & ~~\textbf{Minimizer} & ~~\textbf{Syncmer} & ~~\textbf{UHS} & \textbf{LCP-2} & \textbf{LCP-3} & \textbf{MinHash-1K} & \textbf{MinHash-5M} \\ \midrule
                K-mer size\dag & 15 & 15 & 13 & 10.47 & 26.58 & 31 & 31 \\
                Window size\ddag & 10 & $-$ & 20 & - & - & - & -\\
                S-mer size & $  -$ & 4 & - & - & - & - & -\\ 
                Total k-mers & 665.1M & 371.1M & 983.5M & 595.7M & 253.4M & 1.1K & 6.9M\\
                Unique k-mers & 92.6M & 58M & 20.6M & 0.5M & 159.7M & 1K & 5M\\
                Avg Distance & 4.69 & 8.40 & 3.16 & 5.23 & 12.30 & 2.78M & 454.42\\
                StdDev Distance & 3.15 & 8.20 & 1.36 & 1.85 & 4.21 & 3.2M & 463.73\\
                \bottomrule
            \end{tabular}
            \begin{tablenotes}
                {\footnotesize
                \item \dag We show the average core length for LCP, and \ddag~\textit{L} for UHS. We also provide the statistics for MinHash using sketch sizes of 1,000 and 5,000,000.
                }
            \end{tablenotes}
        \end{threeparttable}
    \end{table*}
}

\newcommand{\TableHgLcpLevels}{
    \begin{table*}[htbp]
        \centering
        \caption{Experimental evaluation of \tool and VG using the HPRC data set on GRCh38}
        \label{tab:hg38-construction}
        \begin{threeparttable}
            \begin{tabular}{@{}LMMMMM@{}}
                \toprule
                \textbf{Metric}     & ~~\tool(4)  & ~~\tool(5)  & ~~\tool(6)  & ~~\tool(7)  & ~~VG \\ \midrule
                Total \# Segments   & 408,484,050 & 196,921,188 & 106,364,013 & 67,566,361  & 490,391,209 \\
                Total \# Links      & 423,677,516 & 212,114,654 & 121,557,479 & 82,759,827  & 519,013,255 \\ 
                \midrule
                Avg. Seg. Len.  (bp)     & 26.18       & 54.32       & 100.56      & 158.31      & 28.23 \\
                StdDev Seg. Len. (bp)    & 11.80       & 31.99       & 80.56       & 184.44      & 9.67 \\\midrule
                Exec. Time (sec)    & 475.98      & 421.04      & 391.49      & 375.14      & 4958.44 \\ 
                Peak memory (GB)    & 8.62        & 6.99        & 6.25        & 5.89        & 114.56 \\ 
                Output r/GFA (GB)   & 44.24/28.11 & 26.61/18.94 & 19.10/15.05 & 15.89/13.40 & -/33.83 \\ \bottomrule
            \end{tabular}
            \begin{tablenotes}
                \footnotesize
                \item We compared graph construction metrics between \tool and VG using the HPRC data set~\cite{liao2023draft} on GRCh38. The table shows the total number of segments, total number of links, execution time, average segment length, standard deviation of segment length, memory usage (RSS), and output size for different LCP levels of \tool and VG. \revise{Graphs are constructed using all chromosomes.}              
            \end{tablenotes}
        \end{threeparttable}
    \end{table*}
}

\newcommand{\TableYeastLcpLevels}{
    \begin{table*}[htbp]
        \centering
        \caption{Yeast pangenome construction using 100 strains with \tool, VG, and VariantStore}
        \label{tab:yeast-construction}
        \begin{threeparttable}
            \begin{tabular}{@{}LMMMMM@{}}
                \toprule
                \textbf{Metric}     &  \tool(5)  & VG  & VariantStore$^\ast$ \\ \midrule
                Total \# Segments   & 252,142 & 437,196  & 73,080 \\
                Total \# Links      &
                276,486 & 461,631 & 97,440\\ 
                \midrule
                Avg. Seg. Len.  (bp)          & 47.97 &  27.67 & 165.51 \\
                StdDev Seg. Len. (bp)          & 31.86 & 10.34 & 10.87 \\ \midrule
                Exec. Time (sec)         & 0.52 & 2.25 & 9.61  \\ 
                Peak memory (MB)        & 34.63 & 84.17 & 100.80 \\ 
                Output r/GFA (MB)    & 14 & 20 & N/A \\ \bottomrule
            \end{tabular}
            \begin{tablenotes}
                \footnotesize
                \item Comparison of graph construction metrics between  \tool (level 5), VG, and VariantStore using 100 yeast strains~\cite{Strope2015} on 
                \textit{S. cerevisiae} assembly R64 (GCA\_000146045).
                $^\ast$ VariantStore values represent the overall values for all chromosomes. 
            \end{tablenotes}
        \end{threeparttable}
    \end{table*}
}

\newcommand{\LcpanThreadingResults}{
    \begin{figure*}[htbp]
    \centering
    \begin{minipage}{\textwidth}
        \centering
        \includegraphics[width=0.61\textwidth]{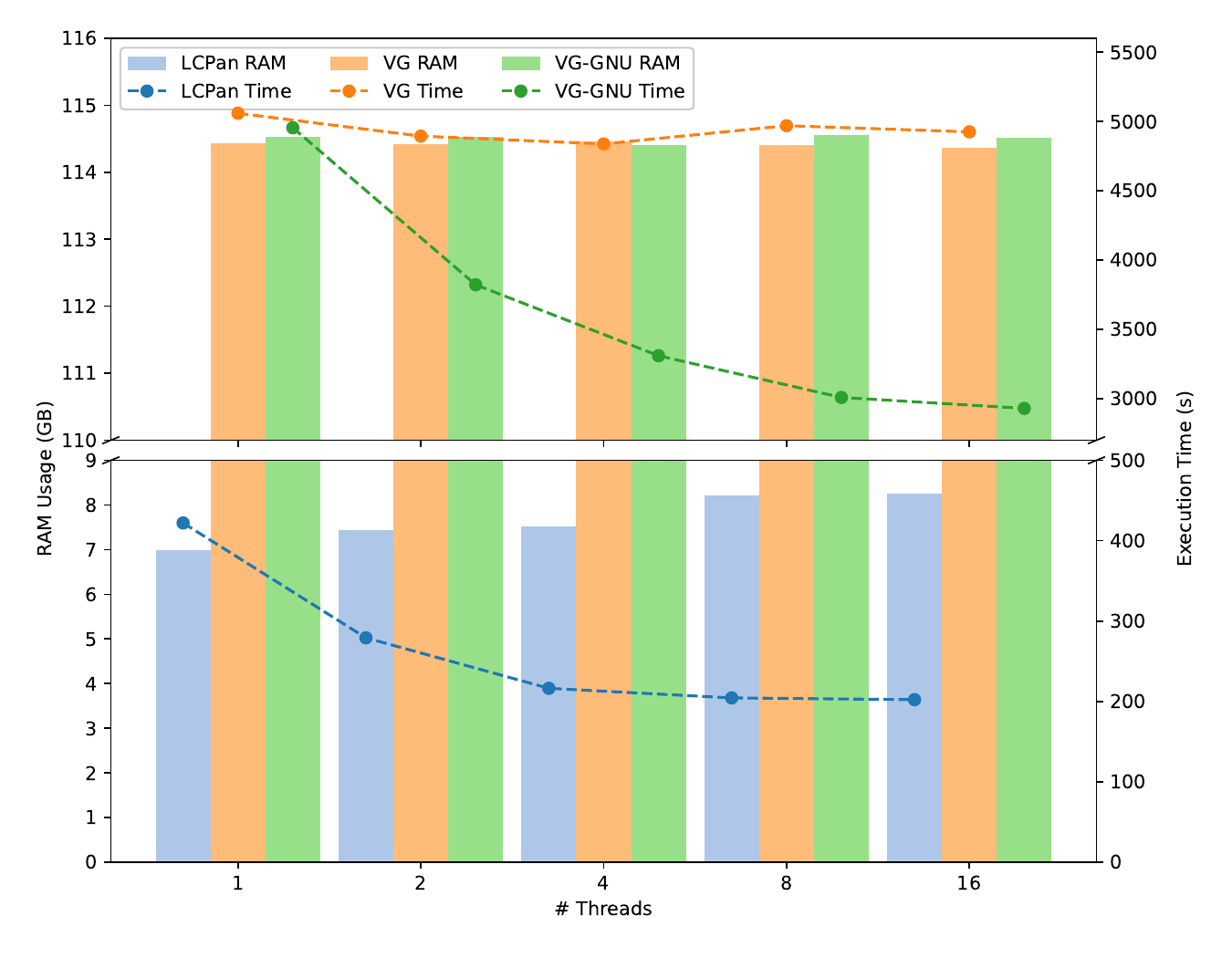} \\
    \end{minipage}
    \caption{\textbf{Multi-thread scaling analysis of \tool and \vgtool using HPRC data on GRCh38.} Memory and run time of graph construction using different numbers of threads. \tool and \vgtool were run using their native multithreading implementations. \vgtool was additionally parallelized using GNU Parallel by distributing single-threaded processes on chunks across concurrent jobs (VG-GNU).}
    \label{fig:lcpan-threading}
    \end{figure*}
}

 \newcommand{\TableAlignmentResultsAll}{
      \begin{table*}[htbp]
         \centering
         \caption{GraphAligner alignment results for PacBio HiFi and ONT reads aligned on \tool and VG graphs.}
         \label{tab:alignment-results-all}
         \begin{threeparttable}
         \resizebox{\textwidth}{!}{%
       \begin{tabular}{@{}BMMMMMM@{}}\toprule
\textbf{Graph}         & \multicolumn{3}{c}{\tool} & \multicolumn{3}{c}{VG}            \\ \midrule 
\textbf{Dataset}       & PacBio HiFi$^\ddag$    & PacBio HiFi$^\dag$         & ONT$^\ddag$   & PacBio HiFi$^\ddag$ & PacBio HiFi$^\dag$       & ONT$^\ddag$ \\ \midrule 
Input Reads            & 820,665                & 5,764,123   & 281,312       & 820,665             & 5,764,123 & 281,312     \\ 
Reads w/ align.        & 820,665                & 5,764,122   & 281,306       & 820,665             & 5,764,123 & 281,308     \\ \midrule
Align. Time (hh:mm:ss) & 06:49:06               & 63:46:26    & 06:19:06      & 08:12:07            & 91:34:26  & 07:42:47    \\
Memory (GB)            & 82.01                  & 220.08      & 114.57        & 83.08               & 411.91    & 103.29      \\ \midrule
Precision              & 0.40                   & 0.49        & 0.41          & 0.38                & 0.38      & 0.41        \\
Recall                 & 0.70                   & 0.53        & 0.70          & 0.72                & 0.72      & 0.71        \\
F1                     & 0.51                   & 0.51        & 0.52          & 0.50                & 0.50      & 0.52        \\ \bottomrule 
\end{tabular}
} 
         \begin{tablenotes}
         \begin{minipage}{\textwidth}
               \footnotesize
                 \item The average lengths of PacBio HiFi and ONT reads are 15,955 bp and 47,000 bp, respectively, corresponding to $\sim$30$\times$ depth of coverage. Precision, recall, and F1 values are comparable across all tests, while \tool-based alignment using \graphaligner is $\sim$1.2$\times$ faster. \revise{Alignments to  $^\ddag$ chromosomes 1, 10, and 22 only, and to $^\dag$ the full genome. Small-scale tests (chromosomes 1, 10, 22) used only those that are known to map to their respective chromosomes, while the full-genome tests used the entire read data sets.}
             \end{minipage}
             \end{tablenotes}
         \end{threeparttable}
\end{table*}
 }

\newcommand{\VisualizeLCP}{
    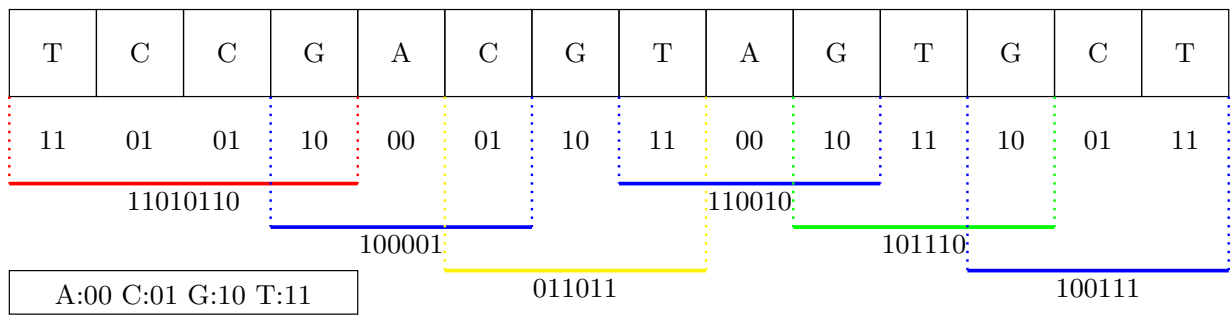
\begin{figure}[ht]
    \centering
    \resizebox{\columnwidth}{!}{
        \begin{tikzpicture}
            \foreach \i in {0,...,13} {
                \draw (\i, 0) rectangle (\i+1, 1); 
            }
            
            \node at (0.5, 0.5) {T};
            \node at (1.5, 0.5) {C};
            \node at (2.5, 0.5) {C};
            \node at (3.5, 0.5) {G};
            \node at (4.5, 0.5) {A};
            \node at (5.5, 0.5) {C};
            \node at (6.5, 0.5) {G};
            \node at (7.5, 0.5) {T};
            \node at (8.5, 0.5) {A};
            \node at (9.5, 0.5) {G};
            \node at (10.5, 0.5) {T};
            \node at (11.5, 0.5) {G};
            \node at (12.5, 0.5) {C};
            \node at (13.5, 0.5) {T};
        
            \node at (0.5, -0.5) {11};
            \node at (1.5, -0.5) {01};
            \node at (2.5, -0.5) {01};
            \node at (3.5, -0.5) {10};
            \node at (4.5, -0.5) {00};
            \node at (5.5, -0.5) {01};
            \node at (6.5, -0.5) {10};
            \node at (7.5, -0.5) {11};
            \node at (8.5, -0.5) {00};
            \node at (9.5, -0.5) {10};
            \node at (10.5, -0.5) {11};
            \node at (11.5, -0.5) {10};
            \node at (12.5, -0.5) {01};
            \node at (13.5, -0.5) {11};
            
            \draw[very thick, red] (0, -1) -- (4, -1);
            \draw[dotted, thick, red] (0, 0) -- (0, -1);
            \draw[dotted, thick, red] (4, 0) -- (4, -1);
            \node at (2, -1.2) {11010110};
        
            \draw[very thick, blue] (3, -1.5) -- (6, -1.5);
            \draw[dotted, thick, blue] (3, 0) -- (3, -1.5);
            \draw[dotted, thick, blue] (6, 0) -- (6, -1.5);
            \node at (4.5, -1.7) {100001};
        
            \draw[very thick, yellow] (5, -2.0) -- (8, -2.0);
            \draw[dotted, thick, yellow] (5, 0) -- (5, -2.0);
            \draw[dotted, thick, yellow] (8, 0) -- (8, -2.0);
            \node at (6.5, -2.2) {011011};
        
            \draw[very thick, blue] (7, -1) -- (10, -1);
            \draw[dotted, thick, blue] (7, 0) -- (7, -1.0);
            \draw[dotted, thick, blue] (10, 0) -- (10, -1.0);
            \node at (8.5, -1.2) {110010};
        
            \draw[very thick, green] (9, -1.5) -- (12, -1.5);
            \draw[dotted, thick, green] (9, 0) -- (9, -1.5);
            \draw[dotted, thick, green] (12, 0) -- (12, -1.5);
            \node at (10.5, -1.7) {101110};
        
            \draw[very thick, blue] (11, -2.0) -- (14, -2.0);
            \draw[dotted, thick, blue] (11, 0) -- (11, -2.0);
            \draw[dotted, thick, blue] (14, 0) -- (14, -2.0);
            \node at (12.5, -2.2) {100111};
        
            \draw (0, -2) rectangle (4, -2.5);
            \node at (2, -2.27) {A:00 C:01 G:10 T:11};
        \end{tikzpicture}
    }
    \caption{\textbf{Processing a string using LCP.} Here, blue underlines the core that satisfies the Local Minimum core, green represents a Local Maximum core,  red corresponds to a Repetitive Interior core, and yellow denotes a Stranded Sequence core.}
    \label{lcp-visual}
    \end{figure}
}

\newcommand{\VisualizeDCT}{
    \begin{figure}[ht]
    \centering
    \resizebox{\columnwidth}{!}{
        \begin{tikzpicture}
            \draw (-0.5, 0) rectangle (1.5, 1);
            \draw (1.5, 0) rectangle (3.5, 1);
            \draw (3.5, 0) rectangle (5.5, 1);
            \draw (5.5, 0) rectangle (7.5, 1);
            \draw (7.5, 0) rectangle (9.5, 1);
            \draw (9.5, 0) rectangle (11.5, 1);
            \draw (11.5, 0) rectangle (13.5, 1);
        
            \node at (0.5, 0.5) {\dots};
            \node at (2.5, 0.5) {100001};
            \node at (4.5, 0.5) {011011};
            \node at (6.5, 0.5) {11101011};
            \node at (8.5, 0.5) {101110};
            \node at (10.5, 0.5) {011011};
            \node at (12.5, 0.5) {\dots};
        
            \draw[dotted, thick, blue] (0.6, -0.1) -- (2.5, -0.7);
            \draw[dotted, thick, blue] (2.5, -0.1) -- (2.5, -0.7);
            \node at (2.5, -1.0) {\ldots};
        
            \draw[dotted, thick, blue] (2.6, -0.1) -- (4.5, -0.7);
            \draw[dotted, thick, blue] (4.5, -0.1) -- (4.5, -0.7);
            \node at (4.5, -1.0) {11};
        
            \draw[dotted, thick, blue] (4.6, -0.1) -- (6.5, -0.7);
            \draw[dotted, thick, blue] (6.5, -0.1) -- (6.5, -0.7);
            \node at (6.5, -1.0) {1000};
        
            \draw[dotted, thick, blue] (6.6, -0.1) -- (8.5, -0.7);
            \draw[dotted, thick, blue] (8.5, -0.1) -- (8.5, -0.7);
            \node at (8.5, -1.0) {00};
            
            \draw[dotted, thick, blue] (8.6, -0.1) -- (10.5, -0.7);
            \draw[dotted, thick, blue] (10.5, -0.1) -- (10.5, -0.7);
            \node at (10.5, -1.0) {01};
        
            \draw[dotted, thick, blue] (10.6, -0.1) -- (12.5, -0.7);
            \draw[dotted, thick, blue] (12.5, -0.1) -- (12.5, -0.7);
            \node at (12.5, -1.0) {\ldots};
        \end{tikzpicture}
    }
    \caption{\textbf{DCT for reducing bitstreams to a new alphabet.} Each block above corresponds to the bitstream of a core. The DCT compares each core's bitstream to its left neighbor to form a shorter alphabet. E.g., the least significant bit of the core bitstream $11101011$, which differs from its left neighbor $011011$ at the fourth index (counting from the right and starting at index 0). The value of this bit is $0$. Therefore, DCT replaces the core bitstream $11101011$ with the concatenation of the bits of $4$, and the value of the differing bit, resulting in $1000$. The figure shows the reduced alphabet inferred by the DCT for each core bitstream, on which LCP can be applied. 
    }
    \label{dct-visual}
    \end{figure}
}

\newcommand{\VisualizeLCPOverview}{
    \begin{figure}[!ht]
    \centering
    \resizebox{\columnwidth}{!}{
        \begin{tikzpicture}
    
            \draw (0, 2) rectangle (4, 2.5);
            \node at (2, 2.27) {A:00 C:01 G:10 T:11};
            
            \foreach \i in {0,...,17} {
                \draw (\i, 0) rectangle (\i+1, 1); 
            }
            
    
            \node at (-0.75, 0.5) {\dots};
            
            \node at (0.5, 0.5) {G};
            \node at (1.5, 0.5) {A};
            \node at (2.5, 0.5) {C};
            \node at (3.5, 0.5) {G};
            \node at (4.5, 0.5) {T};
            \node at (5.5, 0.5) {C};
            \node at (6.5, 0.5) {C};
            \node at (7.5, 0.5) {G};
            \node at (8.5, 0.5) {A};
            \node at (9.5, 0.5) {C};
            \node at (10.5, 0.5) {G};
            \node at (11.5, 0.5) {T};
            \node at (12.5, 0.5) {C};
            \node at (13.5, 0.5) {G};
            \node at (14.5, 0.5) {T};
            \node at (15.5, 0.5) {G};
            \node at (16.5, 0.5) {C};
            \node at (17.5, 0.5) {T};
    
            \node at (0.5, -0.5) {10};
            \node at (1.5, -0.5) {00};
            \node at (2.5, -0.5) {01};
            \node at (3.5, -0.5) {10};
            \node at (4.5, -0.5) {11};
            \node at (5.5, -0.5) {01};
            \node at (6.5, -0.5) {01};
            \node at (7.5, -0.5) {10};
            \node at (8.5, -0.5) {00};
            \node at (9.5, -0.5) {01};
            \node at (10.5, -0.5) {10};
            \node at (11.5, -0.5) {11};
            \node at (12.5, -0.5) {01};
            \node at (13.5, -0.5) {10};
            \node at (14.5, -0.5) {11};
            \node at (15.5, -0.5) {10};
            \node at (16.5, -0.5) {01};
            \node at (17.5, -0.5) {11};
            
            \node at (0, -1.2) {\dots};
            
            \node[blue] at (1.5, -1.2) {100001};
    
            \node[green] at (4.5, -1.2) {101101};
            
            \node[red] at (6, -1.2) {11010110};
            \node[blue] at (8.5, -1.2) {100001};
        
            \node[darkyellow] at (10.5, -1.2) {011011};
        
            \node[blue] at (12.5, -1.2) {110110};
        
            \node[green] at (14.5, -1.2) {101110};
        
            \node[blue] at (16.5, -1.2) {100111};
    
            \filldraw[fill=blue!50, draw=blue, opacity=0.3] (1.5,0.5) ellipse [x radius=1.75, y radius=0.75];
    
            \filldraw[fill=green!50, draw=green, opacity=0.3] (4.5,0.5) ellipse [x radius=1.75, y radius=0.75];
    
            \filldraw[fill=red!50, draw=red, opacity=0.3] (6,0.5) ellipse [x radius=2.25, y radius=0.75];
    
            \filldraw[fill=blue!50, draw=blue, opacity=0.3] (8.5,0.5) ellipse [x radius=1.75, y radius=0.75];
    
            \filldraw[fill=yellow!50, draw=yellow, opacity=0.3] (10.5,0.5) ellipse [x radius=1.75, y radius=0.75];
    
            \filldraw[fill=blue!50, draw=blue, opacity=0.3] (12.5,0.5) ellipse [x radius=1.75, y radius=0.75];
    
            \filldraw[fill=green!50, draw=green, opacity=0.3] (14.5,0.5) ellipse [x radius=1.75, y radius=0.75];
    
            \filldraw[fill=blue!50, draw=blue, opacity=0.3] (16.5,0.5) ellipse [x radius=1.75, y radius=0.75];
    
            \draw[->, thick, bend left=30, dotted] (4.4,-1.5) to (1.5,-1.5);
            \draw[->, thick, bend left=45, dotted] (5.9,-1.5) to (4.5,-1.5);
            \draw[->, thick, bend left=45, dotted] (8.4,-1.5) to (6,-1.5);
            \draw[->, thick, bend left=45, dotted] (10.4,-1.5) to (8.5,-1.5);
            \draw[->, thick, bend left=45, dotted] (12.4,-1.5) to (10.5,-1.5);
            \draw[->, thick, bend left=45, dotted] (14.4,-1.5) to (12.5,-1.5);
            \draw[->, thick, bend left=45, dotted] (16.4,-1.5) to (14.5,-1.5);
    
            \node at (4.5, -2.3) {101};
            \node at (6, -2.3) {00};
            \node at (8.5, -2.3) {01};
            \node at (10.5, -2.3) {11};
            \node at (12.5, -2.3) {00};
            \node at (14.5, -2.3) {111};  
            \node at (16.5, -2.3) {01};
    
            \node at (1, -2.3) {\dots};
            
            \filldraw[fill=blue!50, draw=blue, opacity=0.3] (6.35,-2.3) ellipse [x radius=2.75, y radius=0.75];
            
            \filldraw[fill=blue!50, draw=blue, opacity=0.3] (12.35,-2.3) ellipse [x radius=2.5, y radius=0.75];
    
            \node[blue] at (6.35, -3.7) {101001};
            
            \node[blue] at (12.35, -3.7) {1100111};
            
            \node at (2.5, -3.7) {\dots};
            
            \draw[decorate, decoration={brace, mirror, amplitude=5pt}] (17.5,-1.4) -- (17.5,-1);
            \node[anchor=west] at (18, -1.2) {Level 1 Cores};
            
            \draw[decorate, decoration={brace, mirror, amplitude=5pt}] (17.5,-3) -- (17.5,-1.5);
            \node[anchor=west] at (18, -2.25) {DCT};
            
            \draw[decorate, decoration={brace, mirror, amplitude=5pt}] (17.5,-3.9) -- (17.5,-3.5);
            \node[anchor=west] at (18, -3.7) {Level 2 Cores};
    
    
            \node[anchor=west] at (0, -5.2) {Labels};
            \draw[decorate, decoration={brace, mirror, amplitude=5pt}] (2,-4.5) -- (2,-5.9);
            \node[anchor=west] at (3, -4.7) {$a_0' = \psi (GAC)$};
            \node[anchor=west] at (3, -5.2) {$a_1' = \psi (GTC)$};
            \node[anchor=west] at (3, -5.7) {$a_2' = \psi (TCCG)$};
            \node[anchor=west] at (6, -4.7) {$a_3' = \psi (CGT)$};
            \node[anchor=west] at (6, -5.2) {$a_4' = \psi (TCG)$};
            \node[anchor=west] at (6, -5.7) {$a_5' = \psi (GTG)$};
            \node[anchor=west] at (9, -4.7) {$a_6' = \psi (GCT)$};
            \node[anchor=west] at (9, -5.2) {$a_0'' = \psi (a_1', a_2', a_0')$};
            \node[anchor=west] at (9, -5.7) {$a_1'' = \psi (a_3', a_4', a_5')$};
            
            \node[anchor=west] at (12, -5.2) {\dots};      
    
        \end{tikzpicture}
    }
    \caption{\textbf{Overview of the LCP with DCT and Labeling Paradigm.}} 
    \label{lcp-overview-visual}
    \end{figure}
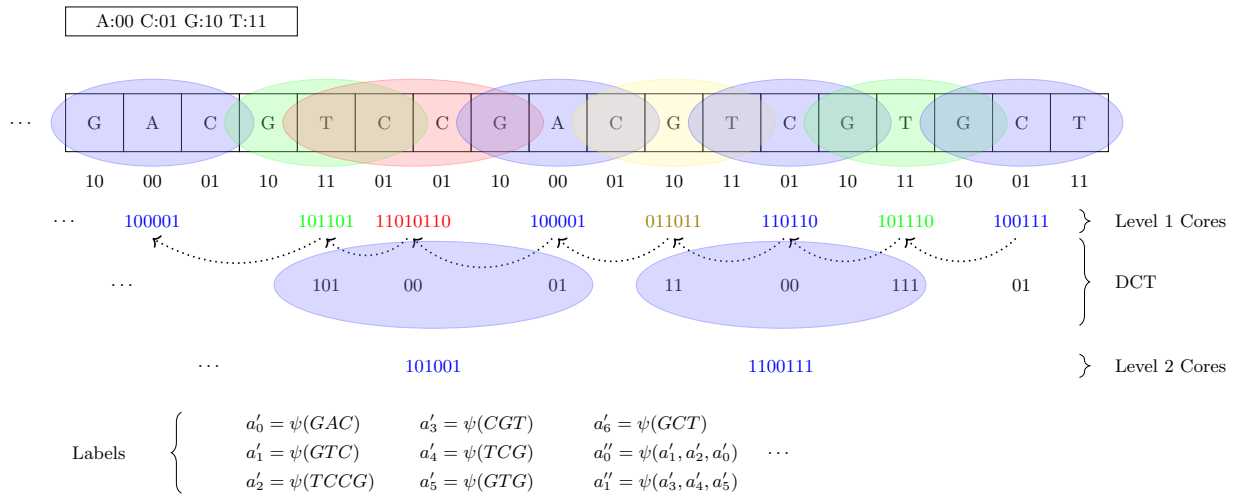
    
}

\newcommand{\VisualLcpanVg}{
    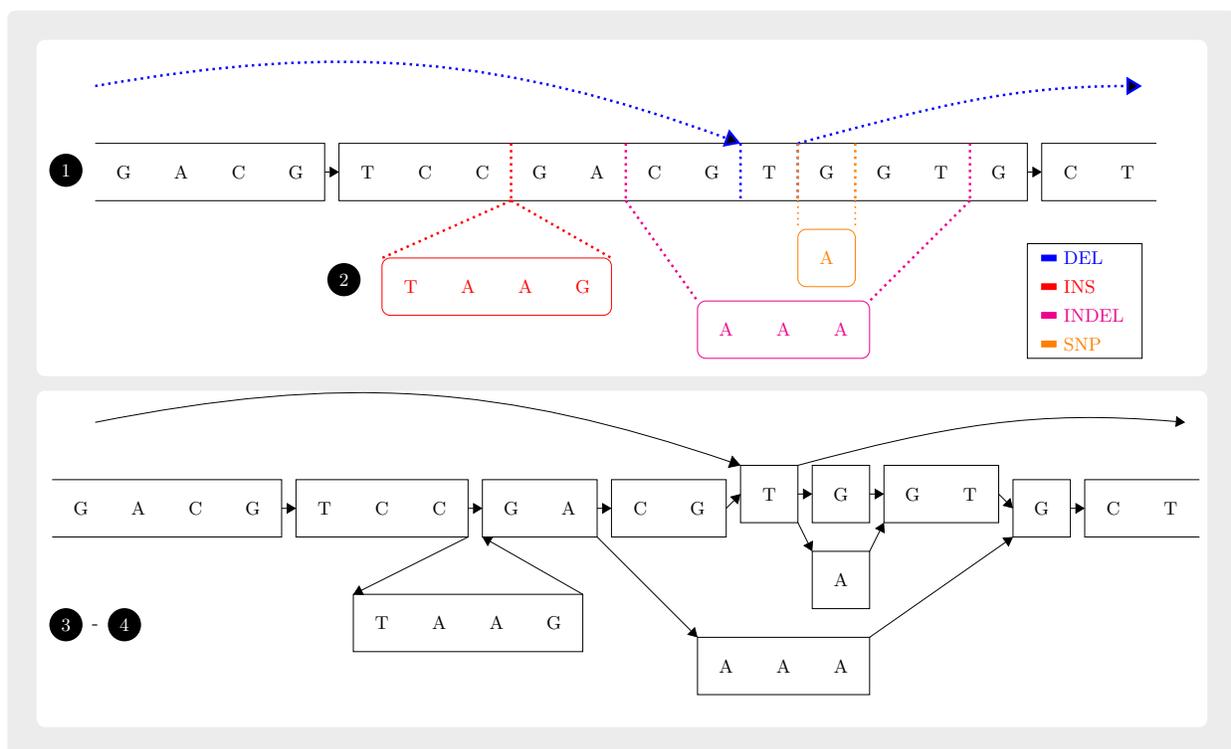
\begin{figure}[!ht]
    \centering
    \resizebox{\columnwidth}{!}{
        \begin{tikzpicture}
        \fill[gray!15, rounded corners] (-1.5, 3) rectangle (19.5, -9.75);
        \fill[white!15, rounded corners] (-1, 2.5) rectangle (19, -3.25);
        \fill[white!15, rounded corners] (-1, -3.5) rectangle (19, -9.25);

        \node[circle, fill=black, text=white, minimum size=0.5cm] at (-0.5,0.27) {1};
        \node[circle, fill=black, text=white, minimum size=0.5cm] at (4.25,-1.6) {2};
        \node[circle, fill=black, text=white, minimum size=0.5cm] at (-0.5,-7.5) {3};
        \node at (0, -7.5) {-};
        \node[circle, fill=black, text=white, minimum size=0.5cm] at (0.5,-7.5) {4};
        
        \begin{scope}[yshift=-0.25cm, scale=0.98] 
            \draw (0,0) -- (4,0) -- (4,1) -- (0,1);
            
            \draw (4.25, 0) rectangle (16.25, 1);
            
            \draw (18.5,0) -- (16.5,0) -- (16.5,1) -- (18.5,1);

            \node at (0.5, 0.5) {G};
            \node at (1.5, 0.5) {A};
            \node at (2.5, 0.5) {C};
            \node at (3.5, 0.5) {G};

            \draw[-{Triangle[fill=black,scale=1.25]}] (4, 0.5) -- (4.25, 0.5);
            
            \node at (4.75, 0.5) {T};
            \node at (5.75, 0.5) {C};
            \node at (6.75, 0.5) {C};
            \node at (7.75, 0.5) {G};
            \node at (8.75, 0.5) {A};
            \node at (9.75, 0.5) {C};
            \node at (10.75, 0.5) {G};
            \node at (11.75, 0.5) {T};
            \node at (12.75, 0.5) {G};
            \node at (13.75, 0.5) {G};
            \node at (14.75, 0.5) {T};
            \node at (15.75, 0.5) {G};

            \draw[-{Triangle[fill=black,scale=1.25]}] (16.25, 0.5) -- (16.5, 0.5);

            \node at (17, 0.5) {C};
            \node at (18, 0.5) {T};
    
            \draw[dotted, very thick, blue] (11.25, 1) -- (11.25, 0);
            \draw[-{Triangle[fill=black,scale=1.25]}, dotted, very thick, blue] (0,2) to[bend left=15, looseness=1] (11.25,1);

            \draw[dotted, very thick, red] (7.25, 1) -- (7.25, 0);
            \draw[dotted, very thick, red] (7.25, 0) -- (5, -1);
            \draw[dotted, very thick, red] (7.25, 0) -- (9, -1);
            \draw[red, rounded corners] (5, -1) rectangle (9, -2);
            \node[red] at (5.5, -1.5) {T};
            \node[red] at (6.5, -1.5) {A};
            \node[red] at (7.5, -1.5) {A};
            \node[red] at (8.5, -1.5) {G};

            \draw[dotted, very thick, magenta] (9.25, 1) -- (9.25, 0);
            \draw[dotted, very thick, magenta] (9.25, 0) -- (10.5, -1.75);
            \draw[dotted, very thick, magenta] (15.25, 1) -- (15.25, 0);
            \draw[dotted, very thick, magenta] (15.25, 0) -- (13.5, -1.75);
            \draw[magenta, rounded corners] (10.5, -1.75) rectangle (13.5, -2.75);
            \node[magenta] at (11, -2.25) {A};
            \node[magenta] at (12, -2.25) {A};
            \node[magenta] at (13, -2.25) {A};

            \draw[dotted, very thick, blue] (12.25, 1) -- (12.25, 0);
            \draw[-{Triangle[fill=black,scale=1.25]}, dotted, very thick, blue] (12.25,1) to[out=15, in=180, looseness=1] (18.25,2);

            \draw[dotted, very thick, orange] (12.25, 1) -- (12.25, 0);
            \draw[dotted, thick, orange] (12.25, 0) -- (12.25, -0.5);
            \draw[dotted, very thick, orange] (13.25, 1) -- (13.25, 0);
            \draw[dotted, thick, orange] (13.25, 0) -- (13.25, -0.5);
            \draw[orange, rounded corners] (12.25, -0.5) rectangle (13.25, -1.5);
            \node[orange] at (12.75, -1) {A};

            \draw (16.25, -0.75) rectangle (18.25, -2.75);
            
            \draw[blue, fill=blue] (16.75, -0.95) rectangle (16.5, -1.05);
            \node[very thick, blue, right] at (16.75, -1) {DEL};

            \draw[red, fill=red] (16.75, -1.45) rectangle (16.5, -1.55);
            \node[very thick, red, right] at (16.75, -1.5) {INS};
            
            \draw[magenta, fill=magenta] (16.75, -1.95) rectangle (16.5, -2.05);
            \node[very thick, magenta, right] at (16.75, -2) {INDEL};

            \draw[orange, fill=orange] (16.75, -2.45) rectangle (16.5, -2.55);
            \node[very thick, orange, right] at (16.75, -2.5) {SNP};
        \end{scope} 

        \begin{scope}[yshift=-6cm, scale=0.98] 

            \draw (-0.75,0) -- (3.25,0) -- (3.25,1) -- (-0.75,1);
            \node at (-0.25, 0.5) {G};
            \node at (0.75, 0.5) {A};
            \node at (1.75, 0.5) {C};
            \node at (2.75, 0.5) {G};

            \draw[-{Triangle[fill=black,scale=1.25]}] (3.25, 0.5) -- (3.5, 0.5);

            \draw[-{Triangle[fill=black,scale=1.5]}] (0,2) to[bend left=15, looseness=1] (11.25,1.25);

            \draw (3.5, 0) rectangle (6.5, 1);
            \node at (4, 0.5) {T};
            \node at (5, 0.5) {C};
            \node at (6, 0.5) {C};

            \draw[-{Triangle[fill=black,scale=1.25]}] (6.5, 0.5) -- (6.75, 0.5);

            \draw[-{Triangle[fill=black,scale=1.25]}] (6.5, 0) -- (4.5, -1);
            \draw[-{Triangle[fill=black,scale=1.25]}] (8.5, -1) -- (6.75, 0);
            \draw (4.5, -1) rectangle (8.5, -2);
            \node at (5, -1.5) {T};
            \node at (6, -1.5) {A};
            \node at (7, -1.5) {A};
            \node at (8, -1.5) {G};
            
            \draw (6.75, 0) rectangle (8.75, 1);
            \node at (7.25, 0.5) {G};
            \node at (8.25, 0.5) {A};

            \draw[-{Triangle[fill=black,scale=1.25]}] (8.75, 0.5) -- (9, 0.5);

            \draw[-{Triangle[fill=black,scale=1.25]}] (8.75, 0) -- (10.5, -1.75);
            \draw[-{Triangle[fill=black,scale=1.25]}] (13.5, -1.75) -- (16, 0);
            \draw (10.5, -1.75) rectangle (13.5, -2.75);
            \node at (11, -2.25) {A};
            \node at (12, -2.25) {A};
            \node at (13, -2.25) {A};

            \draw (9, 0) rectangle (11, 1);
            \node at (9.5, 0.5) {C};
            \node at (10.5, 0.5) {G};

            \draw[-{Triangle[fill=black,scale=1.25]}] (11, 0.5) -- (11.25, 0.75);

            \draw (11.25, 0.25) rectangle (12.25, 1.25);
            \node at (11.75, 0.75) {T};

            \draw[-{Triangle[fill=black,scale=1.25]}] (12.25,1.25) to[out=15, in=175, looseness=1] (19,2);

            \draw[-{Triangle[fill=black,scale=1.25]}] (12.25, 0.25) -- (12.5, -0.25);
            \draw[-{Triangle[fill=black,scale=1.25]}] (13.5, -0.25) -- (13.75, 0.25);
            \draw (12.5, -0.25) rectangle (13.5, -1.25);
            \node at (13, -0.75) {A};

            \draw[-{Triangle[fill=black,scale=1.25]}] (12.25, 0.75) -- (12.5, 0.75);
            
            \draw (12.5, 0.25) rectangle (13.5, 1.25);
            \node at (13, 0.75) {G};

            \draw[-{Triangle[fill=black,scale=1.25]}] (13.5, 0.75) -- (13.75, 0.75);
            
            \draw (13.75, 0.25) rectangle (15.75, 1.25);
            \node at (14.25, 0.75) {G};
            \node at (15.25, 0.75) {T};

            \draw[-{Triangle[fill=black,scale=1.25]}] (15.75, 0.75) -- (16, 0.5);

            \draw (16, 0) rectangle (17, 1);
            \node at (16.5, 0.5) {G};

            \draw[-{Triangle[fill=black,scale=1.25]}] (17, 0.5) -- (17.25, 0.5);

            \draw (19.25,0) -- (17.25,0) -- (17.25,1) -- (19.25,1);
            \node at (17.75, 0.5) {C};
            \node at (18.75, 0.5) {T};
            
        \end{scope}
       
        \end{tikzpicture}
    }
    \caption{\textbf{Overview of the \tool Method.}} 
    \label{lcpan-visual}
    \end{figure}
    
}

\SetKwComment{Comment}{/* }{ */}
\SetKwRepeat{Do}{do}{while}%

\title{\tool: efficient variation graph construction using Locally Consistent Parsing}
%
%
\author{Akmuhammet Ashyralyyev\inst{1} \and
Zülal Bingöl\inst{1} \and
Begüm Filiz Öz\inst{1} \and
Kaiyuan Zhu\inst{2} \and
Salem Malikic\inst{3} \and \\
Uzi Vishkin\inst{4} \and 
S. Cenk Sahinalp\inst{3} \and
Can Alkan\inst{1}}

\authorrunning{A. Ashyralyyev et al.}
%
\institute{Dept of Computer Engineering, Bilkent University, Ankara 06800, Turkey\\
\email{akmuhammet@bilkent.edu.tr, zulal.bingol@bilkent.edu.tr,  filiz.oz@ug.bilkent.edu.tr, calkan@cs.bilkent.edu.tr} 
\and
Department of Computer Science and Engineering, University of California San Diego, La Jolla, CA, USA, 92093\\
\email{kaiyuan-zhu@ucsd.edu}
\and
Cancer Data Science Laboratory, Center for Cancer Research, National Cancer Institute, National Institutes of Health, Bethesda, MD 20892, United States\\
\email{salem.malikic@nih.gov, cenk.sahinalp@nih.gov}
\and 
  University of Maryland Institute for Advanced Computer Studies (UMIACS), College Park, MD 20742 United States \\ 
\email{vishkin@umd.edu}
}

\maketitle

\begin{abstract}

Efficient and consistent string processing is critical in the exponentially growing genomic data era. Locally Consistent Parsing (LCP) addresses this need by partitioning an input genome string into short, exactly matching substrings (e.g., ``cores''), ensuring consistency across partitions. Labeling the cores of an input string consistently not only provides a compact representation of the input but also enables the reapplication of LCP to refine the cores over multiple iterations, providing a progressively longer and more informative set of substrings for downstream analyses.

We present the first iterative implementation of LCP with \api and demonstrate its effectiveness in identifying cores with minimal collisions. Experimental results show that the number of cores at the \(i^{th}\) iteration is \(O(n/c^i)\) for \(c \approx 2.34\), while the average length and the average distance between consecutive cores are \(O(c^i)\). 
Compared to the popular sketching techniques, LCP produces significantly fewer cores, enabling a more compact representation and faster analyses. 
To demonstrate the advantages of LCP in genomic string processing in terms of computation and memory efficiency, we also introduce \tool, an efficient variation graph constructor. We show that \tool generates variation graphs $>$10$\times$ faster than \vgtool, while using $>$13$\times$ less memory. 

\keywords{Locally consistent parsing, genome representation, variation graph}

\end{abstract}

\section{Background}
\label{sec:introduction}

Advances in genomic sequencing technologies and reductions in sequencing costs have significantly increased the production of sequencing data over the last two decades. 
These advancements have boosted the development of a plethora of computational methods for various use cases, including building genome assemblies~\cite{Steinberg2017}, pangenome graph construction~\cite{Pangenome2018}, phylogenetic analysis~\cite{Kapli2020}, and metagenome classification~\cite{Pinto2024}. 

Due to the immense size of genomic sequence data, many of the computational approaches for various tasks such as genome assembly~\cite{Steinberg2017}, metagenome distance estimation~\cite{Ondov2016}, read mapping~\cite{Li2013}, copy number variation~(CNV) genotyping~\cite{Shen2020}, and large-scale sequence database search~\cite{Solomon2016} rely on representing, storing and indexing data as contiguous segments/substrings of length $k$ (i.e., \textit{k-mers}).
However, indexing all possible k-mers uses excessive memory resources. Alternatively, compressed (e.g., Burrows-Wheeler Transformation~\cite{Burrows1994} and FM-index~\cite{Ferragina2000}) or hash-based set membership data structures (e.g., Bloom filters~\cite{Bloom1970}) can be used for smaller memory footprints with the cost of slower lookups~\cite{Marchet2021}.

To reduce memory requirements while keeping the advantages of constant time querying, Roberts et al. proposed to build a \textit{sketch} of the data, and index only a subset of k-mers, called \textit{minimizers}~\cite{Roberts2004}, that can be broadly classified into two classes. 
The simpler class, called \textit{universal minimizers}, includes only k-mers with a hash value less than a specified threshold~\cite{DeBlasio2019}, which are used, for example, in the {\tt rust-mdbg} tool to quickly build minimizer-space de Bruijn graphs~\cite{Ekim2021}. 
The second method finds the k-mers with the minimum hash values within all substrings of size $w$ (i.e., \textit{window}) over the input string. 
\textit{Minimap2}, a state-of-the-art sequence alignment tool, uses this window-based approach to efficiently map long reads to the reference genomes~\cite{Li2018}.
An alternative approach to sampling substrings is Universal Hitting Set (UHS)~\cite{Orenstein2016}, a set of k-mers included in every sequence of length $L$. 
The main drawback of this approach is that computing a minimal set for a given $k$ and $L$ is NP-hard~\cite{Orenstein2016}, and thus approximation algorithms and heuristic methods are employed instead~\cite{Orenstein2016,Ekim2020}. 

In order to address accuracy loss in the presence of sequencing errors, \textit{strobemers}~\cite{Sahlin2021} were introduced to concatenate selected minimizers, thereby improving indexing and error tolerance.
Alternatively, \textit{syncmers}~\cite{Edgar2021} select k-mers based on specific subsequence properties, offering better conservation, lower sampling density, and higher error tolerance compared to classical minimizers.

As an alternative to the aforementioned sketching techniques, Locally Consistent Parsing (i.e., LCP)~\cite{Sahinalp1994,Sahinalp1994a} proposes a string partitioning method with consistent data reduction for building dictionaries like indexes. Sahinalp and Vishkin introduced LCP based on the Deterministic Coin Tossing (DCT)~\cite{Cole1986} technique, 
to construct suffix trees in parallel efficiently. 
LCP was also applied to pattern matching~\cite{Sahinalp1994,Sahinalp1996} and edit distance computation~\cite{Batu2006}. 
A practical but non-iterative variant of LCP was implemented to boost the Lempel-Ziv algorithm for compressing biological sequences~\cite{Hach2012}. Later, locally consistent grammar-based text compression was proposed in ~\cite{diaz2024efficient}.
Other applications of LCP, such as approximate string comparison and text indexing, can be found in ~\cite{Batu2005,Birenzwige2020,diaz2021lms}. \revise{However, except for the non-iterative variant for sequence compression~\cite{Hach2012}, none of the previously proposed algorithms aim to process biological sequences, and only two open-source implementations are currently available~\cite{diaz2024efficient,diaz2021lms}. We note that these implementations are specifically for text indexing and compression, and they do not provide support for utilizing LCP for various use cases.}

\revise{Locally Consistent Parsing (LCP) is a string processing technique that partitions and labels strings into nearly equal-length substrings, known as cores~\cite{Sahinalp1994,Sahinalp1994a}. We note that LCP is not a sketching method; however, the cores can potentially serve as an alternative to sketching. Unlike sketching techniques, LCP ensures that (1) the cores have a uniform positional distribution over the input string, (2) identical cores share the same labels, and (3) each character in the input string is included in at least one core.} In this article, we introduce a generalized, iterative, and practical implementation of LCP, called \api, providing an experimental analysis of the properties of the cores compared to minimizers, syncmers,
MinHash sketches and UHS.
Our analysis highlights that
LCP cores are fewer in number and exhibit longer but more
consistent distribution throughout the processed string.  We
also demonstrate the utility of LCP for variation graph construction using our novel algorithm \tool that outperforms \vgtool, the state-of-the-art variation graph constructor, in both running time ($>10\times$) and memory usage ($>13\times$), while demonstrating slightly better alignment accuracy using \graphaligner~\cite{Rautiainen2020} on HiFi data. We note that both \tool and \vgtool generate variation graphs using a linear reference genome and a VCF file that contains genomic variation, where alternatives like pggb~\cite{Garrison2024} and Minigraph-Cactus~\cite{Hickey2023} build variation graphs using fully assembled contigs and scaffolds. We therefore limit our comparisons to only using \vgtool for data consistency.

\section{Results}
\label{sec:results}

In this section, we provide our analysis of the properties of LCP cores at different levels, as well as \tool efficiency. For our experiments, we used a server with an Intel Xeon CPU (2.2 GHz) with 112 cores and 768 GB of main memory.

\subsection{Properties of LCP Cores}
\label{sec:lcpproperty}
We analyzed the properties of the cores on the first fully assembled human genome (CHM13v2.0) generated by the Telomere-to-Telomere Consortium (T2T)~\cite{Nurk2022}. 
We measured several metrics at each level up to level-8: 1) the number of identified cores, 2) the number of unique core labels, 3) the distances between the starting locations of cores, 4) the lengths of the underlying strings represented by the cores, and 5) the execution time to partition the input string up to the specified LCP level. 
Table~\ref{tab:lcp-fasta-dct1} summarizes our findings and demonstrates that the entire genome can be represented by only 3.6 million cores at level 8, and each level nearly halves the number of cores.

As shown in Table~\ref{tab:lcp-fasta-dct1}, the total number of cores decreases progressively as the levels increase, from approximately 1.4 billion at level-1 to only 3.6 million at level-8. 
Between each successive level, the core count is reduced by a factor of 0.43.
Because the reduction is consistent across successive levels, we can estimate the number of cores in each level for any given input string.
This, in turn, enables us to calculate the memory footprint and the computational overhead for many genomic data applications.
Conversely, the average distance between the starting coordinates of subsequent cores is 2.25 bp in level-1, and it reaches 854.87 bp in level-8. This change corresponds to a fold increase ranging from approximately 2.25 to 2.34 between levels.

\TableLcpCHM

\revise{
This observation empirically supports the theoretical analysis that $2 \leq c \leq 3$, as originally developed by Sahinalp and Vishkin~\cite{Sahinalp1994a}. We note that no k-mer-based sketching scheme (e.g., minimizers and syncmers) can guarantee fixed distances between consecutive substrings. Minimizers are only guaranteed to have a maximal gap of size $w$ between them. We refer the reader to other work in the literature for an analysis of minimizer densities~\cite{Ingels2025}.}

Next, we examined the average length of the sequences represented by cores at each level. At the first level, the average core length is 3.67, aligning with expectations since LMIN, LMAX, and SSEQ cores (given an alphabet of four letters) each span only three characters, whereas RINT cores are longer. The core lengths increase steadily at higher levels, exceeding 1.9 kb by level 8.

We then compared the LCP method against commonly used sketching methods, namely minimizers, syncmers, UHS, and MinHash, using the same input genome to assess the performance of each method regarding the number of selected k-mers and the distances between each selected k-mer.
The LCP levels were selected as 2 and 3 because the average length of the cores is similar to the commonly used k-mer sizes in practice.
We selected $k=15$ and $w=10$ for the canonical minimizers as these are the default values in the popular minimizer-based alignment tool, Minimap2~\cite{Li2018}. 
For the syncmer analysis, we selected $k = 15$, $s=4$, and $t=0$, allowing s-mers to be located only at the beginning of the k-mer, based on the recommended settings in the original publication~\cite{Edgar2021}. 
\revise{For the UHS analysis, we downloaded the pre-computed k-mer sets released by the authors for k=13 and L=20. Similarly, for MinHash, we analyzed the default Mash sketch size of 1,000 k-mers, as well as 5 million k-mers (k=31), which roughly corresponds to the number of cores in LCP level 8. We then mapped these k-mers to the CHM13v2.0 assembly and retained only those that matched exactly. Note that because of the repetitions in the genome, the number of observed k-mers in the genome is slightly higher than the k-mer counts in UHS and MinHash sketches.}

As shown in Table~\ref{tab:min-sync-uhs-fasta}, for $k=15$, we found over 665 million minimizers in CHM13v2.0, among which 92.6 million were unique. 
The syncmer count was smaller, at 371 million, with 58 million unique users. 
Since UHS are computed independently from a reference genome, the unique count was smaller, at 20.6 million. 
Still, when mapped to the genome, it returned 983.5 million locations, the highest value among all techniques tested. 
This is likely because the UHS set contained many k-mers with long homopolymers. 
MinHash is the coarsest-grained sketching method in our tests. 
As expected, it showed the smallest number of k-mers, which is a user-defined parameter. 
Similar to our LCP analysis, we calculated the average distances between each pair of minimizers, syncmers, UHS, and MinHash. 
On average, the distance between two consecutive minimizers was 4.69 nucleotides for $k=15$, whereas the distance between two consecutive syncmers was 8.40 nucleotides. 
UHS k-mers were more densely organized in the genome, where the average distance was 3.16 nucleotides. Once again, MinHash demonstrated the widest range of distances between selected k-mers, with 454.42 bp average distance for the large sketch size (5 million) and 2.78 Mb distance for the default sketch size of 1,000 in Mash.
The cores in LCP level-2 were, on average, 10.47 bp, and in level-3, were, on average, 26.58 bp; the former is slightly shorter, and the latter is slightly longer than the minimizer and syncmer lengths we considered.
Although shorter, the number of level-2 cores was less than that of minimizers for $k=15$. 
Importantly, the number of unique level-2 cores is several orders of magnitude smaller than the minimizers or syncmers of any size, thanks to consistent identification of cores using LCP. 
The number of level-3 cores is less than all the alternatives we considered, except MinHash. 
Perhaps most importantly, among all the other sketches, LCP deviates less proportionally from the average distances.
Higher LCP levels show even lower numbers of cores, therefore reducing the computational cost for storage, indexing, and query (Table~\ref{tab:lcp-fasta-dct1}).

\TableMinSyncUhsCHM

We note that finding minimizers and syncmers requires only a linear scan of the input string and simple lexicographic comparison, therefore faster than identifying cores in LCP in practice. 
Computing \revise{a minimum-size} UHS, on the other hand, is NP-hard~\cite{Orenstein2016}; thus, the UHS k-mer sets can only be approximated. 
However, the granularity of sketches generated by LCP is adjustable using different levels. 
Therefore, depending on the specific use case, longer but fewer or shorter but more frequent cores can be used.
On the other hand, higher LCP levels are constructed by parsing lower levels; therefore, cores at several levels can be computed in one pass of the input string. 
Similar flexibility by minimizers and syncmers can be achieved only by changing the parameters $k$, $w$, and $s$, which will require multiple passes on the input string and incur significant bottlenecks on both the sketching method's run time and memory footprint. However, we leave a rigorous comparison of LCP  with sketching methods for read mapping as future work.

\subsection{Properties of \tool Graphs and Construction Performance}
\label{sec:results-stats}

\TableHgLcpLevels

\TableYeastLcpLevels

We constructed variation graphs, with varying LCP levels, using \revise{all chromosomes of the}  human reference genome (GRCh38) as the backbone and the genomic variations released by the Human Pangenome Reference Consortium~\cite{liao2023draft} \revise{(Table~\ref{tab:hg38-construction})}.
We first evaluated the performance of \tool at different LCP levels. 
We observed a consistent decrease in the number of segments in \tool, from 408 million at level-4 to 67 million at level-7. 
The number of links followed a similar pattern, decreasing from 423 million to 82 million between levels 4 and 7. 
The average segment length increased from 26.18 bp to 158.31 bp as the LCP level increased, while the standard deviation of the segment lengths rose from 11.80 bp to 184.44 bp. 
Reducing the number of segments and links helped decrease the memory usage from 8.63 GB to 5.89 GB.
We also observed a similar decreasing trend in the final output size in GFA and rGFA formats up to level-7.
\vgtool generated a variation graph with an average segment length of 28.23 bp, comparable to LCP level-4, and a standard deviation of 9.67 bp. However, it used significantly more memory at 114.56 GB.
\tool was $>$10$\times$ faster than \vgtool at level-4 while using 13.29$\times$ less memory, with additional improvements at higher LCP levels.
Note that \vgtool cannot process large inputs, particularly for chromosomes that contain tens of millions of base pairs\footnote{\href{https://github.com/vgteam/vg/issues/4152}{https://github.com/vgteam/vg/issues/4152}} and with VCF files that contain millions of variations. 
To manage such large data sets, we followed the \vgtool user manual to process the genome in segments and then merged the output using the \textit{vg combine} command.
The construction of variation subgraphs was completed in 35 minutes using 9.18 GB of memory, while merging required an additional 47.5 minutes and 114.56 GB of memory. 
In addition to the construction time, \vgtool requires an additional $\sim$11 minutes and 91 GB of memory to convert the output graph to GFA format. 

\revise{As an alternative to the human genome, we repeated the graph construction analysis using 100 yeast strains~\cite{Strope2015} (Table~\ref{tab:yeast-construction}). We built variation graphs with \tool, \vgtool, and VariantStore~\cite{Pandey2021}. We note that VariantStore does not generate a mappable variation graph; rather, it is primarily designed to index genomic variants from multiple genomes. Our observations were in line with the human genome experiment, albeit to a lesser degree, where \tool showed a 4.33$\times$ speedup over \vgtool and an 18.5$\times$ speedup over VariantStore, with improvements of 2.43$\times$ and 2.91$\times$ in peak memory usage, respectively.}

\subsection{Scaling to multiple threads}
\label{sec:results-threads}

\LcpanThreadingResults

\tool first processes the reference genome to build the backbone graph sequentially and then assigns the set of variations to multiple threads to amend the graph. 
The partition consistency provided by LCP enables \tool to efficiently scale computation into multiple threads (\revise{see Section~\ref{sec:lcpan-vg}} for details). 
To evaluate the parallelization performance, we constructed the same variation graph using different numbers of threads \revise{for both \tool and \vgtool} (Figure~\ref{fig:lcpan-threading}). \revise{We note that \vgtool documentation recommends parallel construction in two ways: 1) a command-line parameter (\texttt{vg construct -t}), and 2) by partitioning the genome and VCF file into chunks and distributing the task across multiple threads using the GNU Parallel tool~\cite{Tange2011a}. We have used both methods for \vgtool in our evaluation.}

We observed that the memory usage was not significantly affected across different thread settings \revise{for both \tool and \vgtool}. 
Meanwhile, \tool speed was improved up to \revise{8} threads, and more threads did not improve the run time.
This plateau in thread scaling is caused by two limiting factors: the sequential processing of the reference genome and the task allocation speed of the main thread, which distributes the processing to the worker threads.

\revise{For \vgtool, we observed that the native command-line parameter had no effect on improving efficiency, whereas the GNU Parallel-based technique provided speedups of up to 16 threads. In comparison, \tool exhibits better scalability in both cases.}

\subsection{Alignment accuracy}
\label{sec:results-alignment}

We next evaluated the accuracy of read mapping to variation graphs generated by \tool and \vgtool, using various read (PacBio HiFi and ONT) and variation graph data sets. 
As outlined in the introduction, a direct comparison with pggb and Minigraph-Cactus is methodologically inconsistent because \tool and \vgtool construct variation graphs by augmenting a linear reference with variants from a VCF file, whereas pggb and Minigraph-Cactus build graphs \textit{de novo} using multiple whole-genome assemblies. We therefore limited our comparisons to only \vgtool.
We performed the experiments with chromosomes 1, 10, and 22 (GRCh38), which provide a range of lengths and repeat/duplication complexity. 

We first constructed both \tool and \vgtool variation graphs using only chromosomes 1, 10, and 22 (GRCh38) and the variants released by the Human Pangenome Reference Consortium (HPRC)~\cite{liao2023draft}.
\vgtool constructed the graph in 6m 18s  using 12.91 GB of memory, while \tool constructed the graph in only 2m 16s  using 2.17 GB of memory.
We set the k-mer size in the \vgtool graph to 64 and the LCP level to 5 for \tool to make the average segment sizes (i.e., lengths of the sequences represented in nodes) comparable. 
The resulting graphs included segments of average lengths 56.25 bp and 55.17 bp for \tool and \vgtool, respectively.

Next, we aligned both PacBio HiFi and ONT reads generated from the HG002 sample by the Telomere-to-Telomere (T2T) Consortium (v1.0 assembly polishing/mapping data)~\cite{Hansen_2025}, to the constructed variation graphs. Since \tool is focused solely on variation graph construction, and \vgtool is optimized for short read data~\cite{Ma2023}, we used \graphaligner~\cite{Rautiainen2020} with 96 threads in our mapping experiments for both types of graphs.

\TableAlignmentResultsAll

We used the structural variation (SV) call sets released by the
the Genome in a Bottle Project~\cite{Zook2020}\footnote{\href{https://ftp-trace.ncbi.nlm.nih.gov/ReferenceSamples/giab/data/AshkenazimTrio/analysis/}{https://ftp-trace.ncbi.nlm.nih.gov/ReferenceSamples/giab/data/AshkenazimTrio/analysis/}} using the \pbsv tool~\cite{pbsv}, as a proxy to calculate alignment accuracies. 
Our evaluation involved calculating precision, recall, and F1-scores, with a tolerance of a $\pm$100 bp margin around the SV breakpoints. To ensure a fair comparison, we also generated SV calls using the \pbsv tool, and restricted the variations to match the ground truth.
Table~\ref{tab:alignment-results-all} summarizes the mapping results.
Nearly all reads were aligned to the \vgtool and \tool graphs across two data sets.
However, we observe that \graphaligner completed alignments on the \tool graph $\sim$1.2$\times$ faster than on the \vgtool graph, though it required slightly more memory on the ONT dataset. Overall, performance metric values for accuracy across different data sets using \tool and \vgtool graphs remained consistent.

\revise{
Finally, we repeated our read mapping test using the full variation graphs we created with all chromosomes in GRCh38. Since \graphaligner requires a substantial amount of time for read mapping, we limited our test to only the PacBio HiFi data sets. In this test, we observed that \graphaligner was 1.4$\times$ faster using the \tool graph compared to the \vgtool graph, with 1.87$\times$ less peak memory (Table~\ref{tab:alignment-results-all}).
}

\section{Discussion}
\label{sec:discussion}

In this paper, we showed that LCP provides a consistent, uniform, and efficient approach for genomic string partitioning. We also showed that LCP  helps improve computational efficiency and parallelization of variation graph construction. 
One of the key features of the LCP technique is its flexibility in selecting levels, allowing the adjustment of the granularity of the cores based on the requirements of specific use cases.  
For example, middle-level (i.e., 4-5) can be more appropriate for genomic data compression, where large CNV discovery could be achieved using higher levels. LCP can also be used in place of k-mers or minimizers in genome assembly using long and accurate reads. The relationships between levels and the length and distance properties at each level could also be utilized; for example, a hierarchical strategy can be used to align long sequences. However, the analysis of ``best practices'' for LCP level selection for each potential use case requires rigorous analysis and experimentation, which we leave as future work.

We also demonstrated that using LCP, our variation graph construction method, \tool, significantly reduces memory consumption while maintaining high computational efficiency. 
Partition consistency supplied by LCP enables better parallelization, more efficient resource utilization, leading to lower memory usage and faster execution times. 
Experimental results show that \tool consistently outperforms \vgtool in execution speed and scales well with an increasing number of threads.

While \tool currently only implements the graph-building step, \vgtool is an all-inclusive tool that builds the graph, aligns reads, and genotypes variants.
Although \vgtool supports multi-threading, its performance did not scale efficiently with an increasing number of threads.
On the other hand, even without indexing and cache optimizations, \tool surpasses \vgtool in graph construction efficiency while reducing memory usage and execution time. 
This demonstrates the inherent efficiency of LCP-based variation graph construction. However, a formal analysis for alignment accuracy remains as future work.

Although originally developed more than 30 years ago, the practical use of Locally Consistent Parsing remained limited. 
LCP applies, in theory, to various problems arising from sequence analysis, such as edit distance approximation~\cite{Batu2005}, string embeddings~\cite{Batu2006}, memory-efficient text indexing~\cite{Birenzwige2020}, and compression~\cite{diaz2024efficient}. 
Here, we provide an API to enable researchers to fully exploit the LCP method for different use cases.
Our implementation can easily be extended for non-DNA alphabets, enabling the processing of, for example, protein sequences. 
Further research using LCP, made practical through \api, will potentially prove the efficiency and accuracy of various algorithms and impact biomedical research. 

\section{Methods}
\label{sec:methods}

The key properties of LCP have been formally defined earlier in~\cite{Sahinalp1994a}:
 
 \begin{definition}
    Partition Consistency: Given an input string $S$, suppose that a substring $X_i$ starts at position $i$ and $X_j$ starts at position $j$. If $X_i=X_j$, then with a possible exception of left and right margins, $X_i$ and $X_j$ are partitioned in the same way.
\end{definition}
\begin{definition}
    Labeling Consistency: All cores that consist of the same characters are assigned the same labels. 
\end{definition}

Note that these properties of LCP supersede the \textit{window guarantee} provided by sketching techniques since all of the input string is represented within cores.

Below, we describe a simple variant of LCP, expressed in terms of four rules, to identify cores in an input string from a given alphabet (e.g., the $4$-letter DNA alphabet). 
We will later show how to apply LCP iteratively using DCT in Section~\ref{sec:prelim2}.
\begin{enumerate}
    \item \textit{Local Minimum Rule (LMIN):} A substring $w, |w|=3$, $w=xyz$, is a core, if the middle character $y$ is a \textit{proper} local minimum (w.r.t. the lexicographic ordering of the characters). 
    Specifically, for $x \neq y$ and $y \neq z$, the middle character $y$ must satisfy $x > y$ and $y < z$.
    
    \item \textit{Local Maximum Rule (LMAX):} A substring $w, |w|=3$, $w=xyz$, is a core, if the middle character $y$ is a proper local maximum, and neither $x$ nor $z$ is a local minimum.  
    Specifically, for $w=xyz$ that appears within the superstring $sxyzt$, the middle character $y$ satisfies $x < y$, $y > z$, and additionally $s \leq x$ and $z \geq t$.
    
    \item \textit{Repetitive Interior Rule (RINT):} A substring $w, |w|>3$ is a core if all its characters except the first and the last characters are identical. 
    Specifically, for $w=xy^iz$, where $i > 1$, $x\neq y$, and $y\neq z$, the substring $w$ satisfies the condition of being a core.
    
    \item \textit{Stranded Sequence Rule (SSEQ):} A substring $w, |w|>2$ is a core if its characters are either strictly increasing or decreasing w.r.t. the lexicographic order, and only the first and last characters are part of either a \textit{LMIN}, \textit{LMAX}, or a \textit{RINT} cores. 
    Specifically, if $w=xyz \textcolor{blue}{a_1 \dots a_n} klm$, where $n\geq1$ and $xyz$ and $klm$ are identified as cores and $z<a_1<\dots<a_n<k$ or $z>a_1>\dots>a_n>k$, then $z \textcolor{blue}{a_1 \dots a_n} k$ is a core (SSEQ type). 
\end{enumerate}

Figure~\ref{lcp-visual} depicts an example of the cores identified in a short DNA sequence. As the cores are identified, the two-bit encoding of the underlying characters (i.e., A = 00, C = 01, G = 10, T = 11) is used to form the core's \textit{bitstream}. Note that we use the term \textit{core alphabet} to refer to the bitstreams that form the cores, specifically the DNA encoding used in the first LCP iteration.

\VisualizeLCP

Given any input string\footnote{From this point on, we will use the terms \textit{string} and \textit{sequence} interchangeably.}, LCP ensures that the distances between consecutive cores are small.
We now show that the cores identified through the rules above satisfy the \textit{Contiguity Property} and the \textit{Adjacency Property}.
 We provide the lemma definitions below and the formal proofs in the Supplementary Note.

\begin{lemma}
\label{contiguity}
Contiguity Property: There are no gaps between any pair of consecutive cores identified by LCP.
\end{lemma}

The contiguity property guarantees that the cores fully represent the input string and all characters in the input are included in at least one core.\footnote{This may be violated by a small number of characters at the extreme ends of the input string.}

\begin{lemma}
\label{adjacency}
Adjacency Property: within a substring of length $3$, at most $2$ characters can be the starting positions of the cores. 
\end{lemma}

The adjacency property dictates that for $w=xyzlmn$, if $xyz$ is a core, the next two cores cannot start simultaneously at $y$ and $z$.
As the number of cores that start in a window of size 3 cannot exceed 2, we can assert that, in the worst case, the total number of cores will be bounded by $2n/3$, where $n$ is the length of the input string. 
The worst case occurs, for example, in $S=gcggcggcg\dots gcg$, where $c < g$.
In this sequence, the number of LMIN cores (e.g., $gcg$) will be $n/3$, and the number of RINT cores (e.g., $cggc$) will be $n/3-1$, resulting in a total of $2n/3-1$ cores.
However, if LMAX and SSEQ cores are also identified in the string, the cores are positioned further apart, leading to a smaller number of cores.

\subsection{Iterative Application of Locally Consistent Parsing}
\label{sec:prelim2}

\revise{
Our goal is to apply LCP \textbf{iteratively}, so that at higher levels, cores become longer, the distance between consecutive core start positions increases, and the total number of cores decreases, all by approximately a constant factor per level. A direct attempt to do this by reapplying the LCP rules to the raw \textbf{bitstream representations} of level‑$i$ cores is problematic: once we move beyond the DNA alphabet, the ``symbols'' (core bitstreams) effectively come from a much larger, variable-length alphabet, and applying the LCP core-identification rules on such an alphabet can lead to \textbf{uncontrolled growth} in spacing and loss of the even spacing properties ensured by the LCP lemmas at the base level.

\VisualizeDCT

To keep iterative parsing well-behaved, we incorporate \textbf{Deterministic Coin Tossing (DCT)} when constructing levels $i>1$. DCT maps each level‑$i$ core bitstream to a shorter code over a reduced alphabet by comparing it to its \textbf{immediate left neighbor} (Figure~\ref{dct-visual}). Concretely, let $b_{j-1}$ and $b_j$ be the (right-aligned) bitstreams of two consecutive level‑$i$ cores. DCT finds the \textbf{least significant position} $t$ (counting from the right, starting at 0) where $b_{j-1}$  and $b_j$ differ, and replaces $b_j$ by a new code that concatenates (i) the binary representation of $t$ and (ii) the value of $b_j$ at position $t$. If the original bitstreams have length $k$, then $t\in {0,\ldots,k-1}$ and the DCT code has length $\lceil\log k\rceil+1$ bits (the index plus one bit for the value). This reduction controls the effective alphabet size of the sequence on which LCP is applied, so that the resulting ``string of reduced symbols'' continues to satisfy the contiguity/adjacency-style constraints needed for iterative LCP.

\VisualizeLCPOverview

After DCT, we apply the standard LCP core rules to the resulting reduced-symbol sequence to obtain \textbf{level‑($i+1$)} cores. If a level‑($i+1$) core $x$ is the concatenation of consecutive level‑$i$ cores $y,z,w,\ldots$ (i.e., $x=yzw\ldots$), then the level‑($i+1$) bitstream for $x$ is defined as the concatenation of the corresponding DCT codes $y',z',w',\ldots$. Because DCT is deterministic and depends only on adjacent core bitstreams, repeated occurrences of the same block of level‑$i$ cores yield the same reduced representation \textbf{up to boundary effects}, which is sufficient to support locally consistent parsing at higher levels.

Equipped with DCT, each LCP level reduces the number of cores while increasing both the \textbf{average core length} and the \textbf{average distance between consecutive cores} by a constant factor $c$. From the adjacency bound at each level, $c$ is at least 3/2 in the worst case; empirically, on the human genome, we observe $c\approx 2.34$ (Table~\ref{tab:lcp-fasta-dct1}). Thus, at level-$i$, we expect on the order of $n/c^i$ cores representing substrings of average length $\Theta(c^i)$, and the average distance between consecutive core start positions is $\Theta(c^i)$. Each level can be computed in time linear in the length of the level’s representation, and because the representation shrinks geometrically across levels, the total time to compute all levels is linear in the original input length.

Our implementation applies \textbf{only one round of DCT} as a pragmatic tradeoff between LCP's guarantees and the efficiency of the ``anchor''. Each additional DCT round requires increasing the \textbf{core-substring length (and thus overlap between consecutive cores)} to preserve the key property that \textbf{core-ness of a substring is context independent} (a core remains a core wherever it occurs). This added overlap between consecutive cores increases redundancy and reduces sparsity. As we have demonstrated, a single round of DCT already yields a stable, well-behaved empirical distribution of distances between consecutive cores on the human genome, so further rounds offer limited benefit relative to their cost. In rare cases, a single round can create a short stranded region not covered by any core substring; however, empirically, such regions are highly infrequent and do not measurably affect the overall distribution of distances between consecutive cores.
}

\subsection{Labeling Paradigm of LCP Cores}
\label{sec:prelim3}

A key concern in the LCP technique is the process of assigning labels to LCP cores.
This step is essential, as the assigned labels represent the underlying string and must be as distinct as possible. 
To achieve this, we follow the following approach to labeling cores. For a core $x$ at level $i>1$, and if it is composed of level-$(i-1)$ cores $y$, $z$, $w,\ldots$, then the label for $x$ is assigned as $x''= \psi (y'', z'', w'', \ldots)$.
Here, $\psi$ is a hash function designed to combine the labels $y''$, $z''$, $w'', \ldots$ to ensure the distinctiveness of the new core label $x''$. 
For labeling level-1 cores, the labels are directly derived from the characters of the underlying string, ensuring that each character or substring is uniquely represented. 

Figure~\ref{lcp-overview-visual} illustrates these high-level concepts of LCP, DCT, and labeling, to provide a comprehensive visual summary of the overall process.

\VisualLcpanVg

\subsection{String Processing with \api}
\label{sec:lcptools}

We have developed \api as a C-based application programming interface (API) that implements the LCP method, which iteratively processes input strings and generates cores at multiple levels. 
\api is primarily designed for processing genomic sequences as it assumes the input strings are generated from the DNA alphabet (i.e., $\Sigma=\{A,C,G,T\}$). 
\api also includes functions to save and load the cores for later usage. To enhance the flexibility and usability of \api, we provide several options to perform LCP, such as custom alphabet encoding.
Finally, labels are assigned to cores using MurmurHash~\cite{Appleby2008} as $\psi$, ensuring efficient and consistent hashing for labeling (Figure~\ref{lcp-overview-visual}).

\subsection{Variation Graph Construction with \tool}
\label{sec:lcpan-vg}

To highlight the efficiency and accuracy gained using the LCP
technique for string processing, we developed \tool, a variation graph construction tool that leverages the \api API to partition strings rapidly. 
The state-of-the-art variation graph construction tool, \vgtool, internally constructs the graph, performs all computations within it, such as simplification, and then serializes the result, which ultimately leads to high resource consumption. 
Unlike \vgtool, \tool facilitates better memory utilization, leading to smaller storage and memory requirements while maintaining near-uniform partitioning. 

\revise{Figure~\ref{lcpan-visual} illustrates \tool's steps for graph construction. Using the selected LCP level, we partition the initial linear genome into LCP cores per chromosome (step \circled{1}). The initial linear genome yields a simple linear “backbone” graph where nodes represent cores, and edges are placed between pairs of consecutive cores of the associated LCP level. For each variation in the input VCF file, we apply the respective operation for the variation (step \circled{2}) to each of the associated sets of LCP cores in this graph. Each branch created between the consecutive LCP cores generates an alternative path in the graph. If the offset of a variation overflows the current LCP core’s range, we reassign it to the next appropriate LCP core. \tool then further partitions the core  (step \circled{3}) to integrate additional variations from alternative haplotypes into the graph (step \circled{4}), which, in its final form, is referred to as the variation graph.} Finally, for both ends of the sequences, we allocate separate segments to represent the reference sequence completely.
Note that there might be at most two such segments in a given ungapped sequence.

\revise{The fundamental strength of \tool lies in the guarantees provided by Locally Consistent Parsing (LCP), particularly regarding partition consistency. This consistency is essential, as it ensures that the partitioned substrings maintain near-uniform lengths. By sequentially processing variation files stored in sorted order, variations within the offsets of LCP cores can be organized and processed concurrently and independently in batches. Moreover, consistent partitioning and labeling facilitate rapid matching, as LCP ensures that sequences are parsed identically and independently, even when separated by megabases. Furthermore, higher LCP levels lead to longer cores and fewer partitions during parsing. Each partition then contains more variations that must be processed together, resulting in more concentrated work per execution unit. Consequently, performance initially improves with the addition of more threads, but after a certain threshold (e.g., 8 threads), the benefit of adding more threads diminishes due to the limits of parallelization.}

\section*{Data availability}
An open source (BSD 3-Clause License) implementation of \api is available at \url{https://github.com/BilkentCompGen/lcptools} and \tool is available at \url{https://github.com/BilkentCompGen/lcpan}.

\section*{Ethics declarations}
\subsection*{Ethics approval and consent to participate}
Not applicable.

\subsection*{Consent for publication}
Not applicable.

\subsection*{Competing interests}
No competing interest is declared.

\section*{Author contributions statement}
A.A., Z.B., K.Z., U.V., S.C.S., and C.A. developed the methods and conceived the experiments. A.A., Z.B., and B.F.O. conducted the experiments. A.A., Z.B., S.M., S.C.S., and C.A. wrote and reviewed the manuscript. 

\section*{Acknowledgments}
We thank Ricardo Román-Brenes for technical support during the implementation phase, Mahmud Sami Aydın for their insightful comments and suggestions, and Ege Şirvan for helpful discussions on the early versions of variation graph construction. 

\section*{Funding}
This work was partially supported by the European Union’s Horizon Programme for Research and Innovation under grant agreement No. 101047160, project BioPIM, and the Intramural Program of the National Cancer Institute.

\clearpage

\clearpage

\appendix
\pagenumbering{Roman}
\setcounter{page}{0}
\setcounter{figure}{0}
\setcounter{table}{0}
\setcounter{lemma}{0}
\setcounter{theorem}{0}
\renewcommand{\thefigure}{S\arabic{figure}}
\renewcommand{\thetable}{S\arabic{table}}

\section*{Supplementary Material}

\subsection*{Proofs}

\subsubsection*{Contiguity and Adjacency Properties}
\begin{lemma}
\label{sup:contiguity}
Contiguity Property: no gaps exist between any pair of consecutive cores identified by LCP.
\end{lemma}
The contiguity property guarantees that cores fully represent the input string and prevents any characters in the input string from not being included in any core.
We prove that LCP satisfies this property below.

\begin{proof}
Given a substring $w=xyzlmn$, one of the following must be correct.

\begin{enumerate}
    \item If $xyz$ is an LMIN core, then the substring $zlm$ or $lmn$ may satisfy the LMAX rule ($y < z < l$ and $l > m \geq n$; or $y < z < l < m$ and $m > n$), assuming $n$ is not the middle character of another LMIN core. 
    On the other hand, if the lexicographic order of subsequent characters increases (i.e., $y < z < l < m < n$), the SSEQ rule will apply after LMIN until the downstream of the input string includes an LMIN, LMAX, or RINT core. 
    Finally, if $z=l$ or $l=m$, a RINT core will follow an LMIN core.
    
    \item If $xyz$ is an LMAX core, then the substring $zlm$ or $lmn$ may satisfy the LMIN rule ($y > z > l$ and $l < m$; or $y > z > l > m$ and $m < n$). 
    Alternatively, if the lexicographic order of subsequent characters decreases (i.e., $y > z > l > m > n$), the SSEQ rule will apply after LMAX until the downstream of the input string includes an LMIN, LMAX, or RINT core. 
    Finally, if $z=l$ or $l=m$; a RINT core will follow an LMAX core.
    
    \item If $x\overline{y}z$ is a RINT core (i.e., $x\overline{y}z=xy^iz, i>1$), then the substring $zlm$ or $lmn$ may satisfy the LMIN rule ($y > z > l$ and $l < m$; $y > z > l > m$ and $m < n$). 
    The RINT substring $x\overline{y}z$ can also be followed by an LMAX core where $zlm$ or $lmn$ may satisfy the LMAX rule ($y < z < l$ and $l > m$; or $y < z < l < m$ and $m > n$). 
    If the lexicographic order of subsequent characters decreases (i.e., $y > z > l > m > n$), or increases (i.e., $y < z < l < m < n$), an SSEQ core will follow the RINT core until one of the LMIN, LMAX, or RINT rules is satisfied. 
    If $z=l$ or $l=m$, the RINT core will be followed by another RINT core.
    
    \item By definition, an SSEQ core shares characters with its neighbor, concluding the correctness of the Contiguity Property.
    
\end{enumerate}
\qed
\end{proof}

We showed in Lemma~\ref{sup:contiguity} that a string can be fully represented without gaps using cores.
Below, we show that there is an upper bound on the number of potential cores for a given string; this has implications for the efficiency of string processing and indexing.
In fact, we demonstrate that the number of cores is a constant factor smaller than the length of the input string after LCP processing\footnote{The number of cores is further reduced after each iterative call to LCP; the cores identified in iteration $i$ are called level $i$ cores.}. 
For that, we define the \textit{Adjacency Property} as follows. 
\begin{lemma}
\label{sup:adjacency}
Adjacency Property: within a substring of length $3$, there could be at most $2$ characters that could be the starting positions of cores. 
\end{lemma}
Similar to Lemma~\ref{sup:contiguity}, the correctness of Lemma~\ref{sup:adjacency} can be proven by analyzing the possible placement of cores in substrings of length three.
\begin{proof}
Given a string $w=xyzlmn$, one of the following must be correct.

\begin{enumerate}
	\item If $xyz$ is an LMIN core, then $zlm$ or $lmn$ may be an LMAX core (i.e., $z\neq l$ and $l\neq m$). 
	By definition, no local minimum can exist adjacent to an LMAX core. 
    Hence, they do not overlap for more than one character (i.e., an LMAX core cannot start with $y$). 
    On the other hand, the closest RINT core after an LMIN core in $w$ can be in $yzlm$, if $z=l$. 
    The next core after $yzlm$ can only start with $l$ since the prefix of no core may be a repeat. 
    Similarly, by definition, an SSEQ core may start only at or after $z$. 
    
	\item If $xyz$ is an LMAX core, the closest LMIN core may only start at $z$. 
	The closest RINT core could start at $y$ if $z=l$, but in that case, no other core may start at $z$ for the same reason outlined above. 
    Again, the same reasoning applies to the next SSEQ core that follows an LMAX core.
    
	\item If $xy^iz$ is a RINT core ($i>1$), the closest possible cores may start after $i$ characters. 
    For example, if $w=xyyznm$, then $xyyz$ is a RINT core, and the next LMIN, LMAX, or SSEQ core may start at $yzn\ldots$.
    
	\item If $xyz$ is an SSEQ core, the closest possible core may start after the third character (i.e., it may be $zlm$). 
 
\end{enumerate}
\end{proof}

\end{document}